\documentstyle[12pt]{article}
\newcommand{\vev}[1]{\left<{#1}\right>}

\newcommand{\tsum}{{\textstyle \sum}}

\def\sectionnumbering {
   \setcounter{equation}{0}
   \renewcommand{\theequation}{\arabic{section}.\arabic{equation}}}

\def\mysection#1{
   \addtocounter{section}{1}
   \setcounter{subsection}{0}
   \sectionnumbering\par\bigskip\par\medskip
   \begin{center}{\sc \arabic{section} \quad  #1 }\end{center}\par}

\def\mysubsection#1{
   \addtocounter{subsection}{1}
   \par\bigskip\noindent
   {\normalsize\it\arabic{section}.\arabic{subsection} \quad #1 }
   \par\medskip}
\begin{document}

\thispagestyle{empty}
\setcounter{page}{0}

\baselineskip 5mm
\renewcommand{\thefootnote}{\fnsymbol{footnote}}
\hfill\vbox{
\hbox{KEK-TH-711}
\hbox{UTHEP-433}
\hbox{YITP-00-51}
\hbox{hep-th/0009107}
}


\baselineskip 0.8cm
\vskip 15mm
\begin{center}
{\large\bf Free Field Approach to String Theory on $\bf AdS_3$}
\end{center}


\vskip 7mm
\baselineskip 0.6cm
\begin{center}
    Kazuo ~Hosomichi\footnote[1]
    {\tt hosomiti@yukawa.kyoto-u.ac.jp},                        \\
       {\it Yukawa Institute for Theoretical Physics}           \\
       {\it Kyoto University, Kyoto 606-8502, Japan}            \\
       \vskip3mm
    Kazumi ~Okuyama\footnote[2]
    {\tt kazumi@post.kek.jp}                                    \\
       {\it High Energy Accelerator Research Organization (KEK)}\\
       {\it Tsukuba, Ibaraki 305-0801, Japan}                   \\
         \vskip1mm
         {\sl and}                                              \\
         \vskip1mm
    Yuji ~Satoh\footnote[3]
    {\tt ysatoh@het.ph.tsukuba.ac.jp}                           \\
       {\it Institute of Physics, University of Tsukuba}        \\
       {\it Tsukuba, Ibaraki 305-8571, Japan}
\end{center}


\vskip 6mm
\baselineskip=3.5ex
\begin{center}{\bf Abstract}\end{center}
\par
\smallskip
 We discuss the correlation functions of the $SL(2,{\bf C})/SU(2)$
 WZW model, or the CFT on the Euclidean $AdS_3$.
 We argue that their calculation is reduced to that of a free theory
 by taking into account the renormalization and integrating out
 a certain zero-mode, which is an analog of the zero-mode integration
 in Liouville theory. 
 Based on the resultant free field picture, we give a simple
 prescription for calculating the correlation functions.
 The known exact two- and three-point functions of generic primary fields
 are correctly obtained, including numerical factors.
 We also obtain some four-point functions of primaries
 by solving the Knizhnik-Zamolodchikov equation,
 and verify that our prescription indeed gives them.

\vskip 5mm

\noindent
{\sl keywords}:~ $SL(2,{\bf C})/SU(2)$ WZW model, free fields, KZ equation,
AdS/CFT correspondence 

\newpage

\renewcommand{\thefootnote}{\arabic{footnote}}
\setcounter{footnote}{0}
\setcounter{section}{0}
\baselineskip = 0.6cm
\pagestyle{plain}

\mysection{Introduction}

   The $SL(2,{\bf R})$ WZW model is the theory of a string
 propagating on $AdS_3$ in the presence of non-zero B-field.
 This theory is closely related to the physics of black holes
 in various dimensions, and it has also an application
 to the AdS/CFT correspondence \cite{Maldacena,Gubser-KP,Witten}.
 At the same time, various conventional techniques of
 conformal field theory allows us to analyze the theory
 in great detail\cite{Giveon-KS,deBoer-ORT,Kutasov-S}.
 Although our knowledge of thie model is far from complete
 due to the difficulty arising from the non-compactness
 of the target space, many recent works\cite{Bars-DM}-\cite{Giribet-N2}
 have clarified some of its fundamental properties.

   For actual applications, the Euclidean version of the
 $SL(2,{\bf R})$ WZW model is equally important.
 The Euclidean $AdS_3$ is given by the quotient space
 $H_3^+=SL(2,{\bf C})/SU(2)$, and the string theory on this space
 is also described by a WZW-like action.
 Since the action reduces to a particularly simple form if we
 use a certain coordinate system, we can analyze the theory from
 the Lagrangian approach \cite{Haba,Gawedzki,Ishibashi-OS}.
 Alternatively, we can analyze the system based on a free field
 realization of current algebra \cite{Bernard-F}-\cite{Saraikin},
 \cite{Giribet-N1,Pesando,Hikida-HS,Giribet-N2}.
 One can also study the theory based on the symmetry and bootstrap
 condition \cite{Teschner1,Teschner2}
 or by solving the Knizhnik-Zamolodchikov(KZ) 
 equation \cite{Kogan-LS}-\cite{Lewis}.
 These different approaches are of complementary use to one another
 and, in order to obtain a complete understanding of this theory,
 it is desirable to clarify how these approaches
 are related to one another.

   In this paper, we discuss the correlation functions
 of the $H_3^+$ WZW model.
 We start from the full Lagrangian and primary fields. Then by 
 integrating out a certain zero-mode
 and taking into account the renormalization, it is argued that the 
 expression of the correlators becomes that of a free theory.
 Using the resultant free field picture,  
 we calculate the correlation functions explicitly and
 obtain the results which are consistent with other approaches.

   This paper is organized as follows.
 In section 2 we briefly summarize the $H_3^+$ WZW model.
 In section 3 we argue how the free field picture emerges,
 and give a prescription for calculating correlators.
 Using this prescription we calculate the two- and three-point
 functions of primary fields, and find that  the known results
 are correctly obtained, including numerical factors.
 In section 4 we obtain some four-point functions of primary fields
 by solving the KZ equation.
 We see that it can be solved explicitly
 if we put a certain condition on the $SL(2,{\bf C})$ spins
 of four primaries.
 Rewriting the solutions in a form manifestly symmetric in four
 vertices, we find that they can be easily reproduced from our
 free field prescription.
 We conclude with a brief discussion in section 5.

\mysection{$H_3^+$ WZW model}
   We begin with a brief review of the $H_3^+$ WZW model. 
   We follow the notations in \cite{Ishibashi-OS} and some details are 
   found there.
   The Euclidean $AdS_3$ is equivalent to the quotient space
 $H_3^+ \equiv SL(2,{\bf C})/SU(2)$,
 and the sigma model on this space with non-zero NS-NS B-field
 is known to be described by a WZW-like action.
 In a certain parameterization the worldsheet action becomes
\begin{equation}
  S = \frac{k}{\pi}\int d^2z
    [\partial\phi\bar{\partial}\phi
    +e^{2\phi}\partial\bar{\gamma}\bar{\partial}\gamma] ~.
\label{S-PG}
\end{equation}
 As in ordinary WZW models, this theory has an affine 
 $SL(2,{\bf C})\times \overline{SL(2,{\bf C})}$ symmetry.
 However, unlike ordinary WZW theories, the left- and
 the right-moving currents are complex conjugate to each other.
 An important class of operators are the primary fields
\begin{equation}
  \Phi_j(z,x) \equiv (e^{\phi}|\gamma-x|^2+e^{-\phi})^{2j} ~,
\end{equation}
 which are characterized by the following OPEs with the 
 $SL(2)$ currents $j^A(z)$:
\begin{equation}
  j^A(z)\Phi_j(w,x)
     \sim -\frac{D^A\Phi_j(w,x)}{z-w} ~,
\label{jPhi}
\end{equation}
\begin{equation}
  D^- =    \partial_x       ~,~~~
  D^3 = x  \partial_x -  j  ~,~~~
  D^+ = x^2\partial_x - 2jx ~,
\label{D-A}
\end{equation}
 and similarly with $\bar{j}^A$.
 Since $j$ merely labels the second Casimir of $SL(2,{\bf C})$,
 there must be a relation between the primary fields of spin
 $j$ and $-j-1$.
 Classically they are related by
\begin{equation}
  \Phi_j(z,x)= \frac{2j+1}{\pi}\int d^2y |x-y|^{4j}\Phi_{-j-1}(z,y),
\end{equation}
 but the coefficient in the right hand side may get
 quantum corrections.

   The action (\ref{S-PG}) has a very simple form and
 allows us to carry out the path integration explicitly.
 In some early works \cite{Haba,Gawedzki} the theory was analyzed
 in the path-integral formalism, using the action (\ref{S-PG})
 and the $SL(2,{\bf C})$-invariant measure
\begin{equation}
 {\cal D}g =
  {\cal D}\phi
  {\cal D}(e^\phi\gamma)
  {\cal D}(e^\phi\bar{\gamma})~.
\end{equation}
 There, it was shown that, after a suitable treatment of divergences
 arising from zero-mode integrals, one can obtain finite result for
 various correlators.
 Based on this formalism, a recent work \cite{Ishibashi-OS} has given
 the two- and three-point functions of primary fields, which agree with
 the results in \cite{Teschner1,Teschner2}.

\mysection{Reduction to free-theory representation}


\mysubsection{General Argument}
To start our argument, we first rewrite the action (\ref{S-PG}) 
by introducing auxiliary
 fields $\beta$ and $\bar{\beta}$, and obtain
\begin{eqnarray}
 S &=& S_{0} + S_{\rm int} ~,  \nonumber \\
 S_{0} &=& \frac{1}{\pi}\int d^2z
   \Big(k\partial\phi\bar{\partial}\phi
    -\beta\bar{\partial}\gamma
    -\bar{\beta}\partial\bar{\gamma}\Big) ~, \label{S-PBG} \\
 S_{\rm int}  &=& -\frac{1}{k\pi}\int d^2z \beta\bar{\beta}e^{-2\phi} ~.
    \nonumber
\end{eqnarray}
The $SL(2)$ invariant  measure is then
\begin{equation}
 {\cal D}\phi{\cal D}(e^{\phi}\gamma){\cal D}(e^{\phi}\bar{\gamma})
       {\cal D}(e^{-\phi}\beta){\cal D}(e^{-\phi}\bar{\beta}) ~.
   \label{fullmeasure}
\end{equation}
Using the above action and measure, correlators are defined by
\begin{equation}
  \vev{ X }
  \equiv 
  \int {\cal D}\phi{\cal D}(e^{\phi}\gamma){\cal D}(e^{\phi}\bar{\gamma})
       {\cal D}(e^{-\phi}\beta){\cal D}(e^{-\phi}\bar{\beta})
  \exp[-S] \cdot X  ~. \label{defcor}
 \end{equation}
 Although the above action seems almost free, it still describes 
 an interacting theory because of the term $S_{\rm int}$.
 In terms of the fields in  (\ref{S-PBG}),  
 the affine symmetry of the original action (\ref{S-PG}) 
is translated into the symmetry under  
\begin{eqnarray}
  \delta \gamma &= &
  \epsilon(\gamma) ~, \quad 
 \delta\bar{\gamma} \ = \ -{1 \over 2} \epsilon''e^{-2\phi} ~, 
  \quad \delta\phi \ = \ -{1 \over 2} \epsilon' ~, \nonumber \\
 \delta\beta &= & -\epsilon'\beta+{k \over 2}e^{2\phi}
 \partial(e^{-2\phi}\epsilon'')  ~,
 \quad \delta\bar{\beta} \ = \ 0 ~. \label{fullsym}
\end{eqnarray}
 Here,
 $\epsilon(\gamma) = \epsilon_-(z) +\epsilon_3(z)
  \gamma+\epsilon_+(z)\gamma^2 $
 and primes denote the derivatives with respect to $\gamma$.
 Similar transformations hold also for $\bar{j}^A$.
 
   In the following we consider the correlation functions
 of the primary fields $\Phi_j$. 
 For later use, we expand $\Phi_j$ in terms of $e^{-2\phi}$:
\begin{equation}
  \Phi_j = \Phi_j^{\rm f} 
  \Bigl( 1 + {\cal O}(e^{-2\phi}) \Bigr) ~, 
\end{equation}
 where 
\begin{equation}
  \Phi^{\rm f}_j(z,x) = |x-\gamma|^{4j}e^{2j\phi} ~. \label{PRIMARY-F}
\end{equation}
 The correlator of $\Phi_j$ is written as
\begin{eqnarray}
  && \vev{ \prod_{a=1}^N \Phi_{j_a}(z_a,x_a) } \label{Xvev} \\ 
  && \qquad =
 \int{\cal D}\phi{\cal D}(e^{\phi}\gamma){\cal D}(e^{\phi}\bar{\gamma})
       {\cal D}(e^{-\phi}\beta){\cal D}(e^{-\phi}\bar{\beta})
   \exp[-S] \prod_{a=1}^N \Phi^{\rm f}_{j_a}(z_a,x_a)
  \Bigl( 1+ {\cal O}(e^{-2\phi})\Bigr)~.  \nonumber
\end{eqnarray}
  From the index theorem for the fields $(\beta,\gamma)$ with spin $(1,0)$,
the zero-mode part of the measure is 
\begin{equation}
  d\phi_0 d\gamma_0 d\bar{\gamma}_0
            d^g\beta_0 d^g\bar{\beta}_0
            e^{2(1-g)\phi_0} ~,
  \label{index}
\end{equation}  
where $\phi_0, \gamma_0, ...$ are the zero-modes of the respective fields and
$g$ is the genus of the worldsheet. 
In this paper, we focus on the case $g=0$, in which 
the above expression becomes
$ d\phi_0d^2\gamma_0e^{2\phi_0}$.

An important point in our argument is that 
one can first perform 
the integration over $\phi_0$. The $\phi_0$ integral in  (\ref{Xvev})  
reads
\begin{eqnarray*}
 &&
  \int d\phi_0 e^{2(\Sigma_a\, j_a +1)\phi_0} 
  \exp\left[e^{-2\phi_0}\int\frac{d^2w}{k\pi}
            \beta\bar{\beta}e^{-2\phi_q}\right] 
   \cdot\Bigl( 1 + {\cal O}(e^{-2\phi}) \Bigr) \\
  &=&
  \frac{1}{2}\Gamma(- \tsum_a j_a -1)
  \left[-\int\frac{d^2w}{k\pi}
        \beta\bar{\beta}e^{-2\phi_q}\right]^{\tsum_a\, j_a +1}
   \cdot\Bigl( 1 + {\cal O}(e^{-2\phi_q}) \Bigr)  ~,
\end{eqnarray*}
 where $\phi_q$ denotes the non-zero mode of $\phi$
and ${\cal O}(e^{-2\phi_q})$ represents 
 the contributions from the higher order
 terms in $e^{-2\phi_q}$. 
 Since the interaction term $S_{\rm int}$ appears
 only in the above form,
 the remaining functional integration reduces to that 
 with respect to the free action $S_{0}$.

Next we perform the functional integration over the non-zero modes.
The functional determinant 
coming from the integration over $\beta\gamma$ and $\bar{\beta}\bar{\gamma}$
gives a shift of the kinetic term of $\phi$:
\begin{equation}
\det{}^{-1}(e^{-\phi_q}\partial e^{2\phi_q}\bar{\partial}e^{-\phi_q}) 
=\exp\left[\int d^2z\left({2\over\pi}\partial\phi\bar{\partial}\phi
+{\phi_q\over 4\pi}\sqrt{g}R\right)\right] . \label{chiralanom}
\end{equation}
The resultant expression for the correlation function of the primaries is
then
\begin{eqnarray}
  && \vev{ \prod_{a=1}^N \Phi_{j_a}(z_a,x_a) } \\ 
  && \quad 
   = \frac{1}{2}\Gamma(- \tsum_a j_a -1) \int d \gamma_0 d \bar{\gamma}_0
    \vev{ \prod_{a=1}^N \Phi^{\rm f}_{j_a}(z_a,x_a)
  \Bigl( 1 + {\cal O}(e^{-2\phi}) \Bigr) 
 S_{\rm int}^{~~\Sigma_a\, j_a + 1}
              }^{\rm f}_{\phi_0=0,\gamma_0,\bar{\gamma}_0} 
   ~, \nonumber 
\end{eqnarray} 
 where the bracket $\vev{A}^{\rm f}_{\phi_0=0,\gamma_0,\bar{\gamma}_0}$
 represents the Wick contraction of $A$ using\footnote{
We omit the terms which arise because of the background metric.
}
\begin{equation}
\begin{array}{rcl}
 \phi(z)  &=& \phi_0   +\phi_q(z), \\
 \gamma(z)&=& \gamma_0 +\gamma_q(z),
\end{array}
~~~
\begin{array}{rcl}
 \vev{\phi_q(z)\phi_q(w)}^{\rm f}_{\phi_0,\gamma_0,\bar{\gamma_0}}
 &=& -b^2\ln|z-w| ~, \\
 \vev{\beta(z)\gamma_q(w)}^{\rm f}_{\phi_0,\gamma_0,\bar{\gamma_0}}
 &=& (z-w)^{-1}~,
\end{array} \label{contraction}
\end{equation}
 with $b^{-2}= k-2$. 
 
   Furthermore, similarly to the discussions in
 \cite{Gawedzki,Ishibashi-OS}, one can show that
 the ${\cal O}(e^{-2\phi})$ terms disappear after the renormalization
 because of the self-contraction of $e^{\phi(z_a)}$s
(at least when the calculation can be carried out).
 Hence we arrive at the expression
\begin{equation}
  \vev{ \prod_{a=1}^N \Phi_{j_a}(z_a,x_a) } 
  = \frac{1}{2}\Gamma(- \tsum_a j_a -1) \int d \gamma_0 d \bar{\gamma}_0
    \vev{ \prod_{a=1}^N \Phi^{\rm f}_{j_a}(z_a,x_a) 
 {S_{\rm int}}^{\Sigma_a j_a + 1}}^{\rm f}_{\phi_0=0,\gamma_0,\bar{\gamma}_0} 
   ~.
 \label{freeint}
\end{equation}
  Since it turns out that $\Phi_j^{\rm f}$
 correspond to the primaries in a free theory, 
 the right-hand side is nothing but a correlation function
 in a free theory with $\gamma_0$-integral. 
 $S_{\rm int}$ plays the role of the screening operator. 

An anomaly term as in (\ref{chiralanom}) may also be obtained 
by changing the measure (\ref{fullmeasure}) to 
\begin{equation}
  {\cal D}\phi{\cal D}\gamma {\cal D}\bar{\gamma}
  {\cal D}\beta{\cal D}\bar{\beta} ~. \label{flatmeasure}
\end{equation}
Such a term and $S_0$ then add up to 
\begin{equation}
 S_{{\rm free}}={1\over\pi}\int d^2z\Big[(k-2)\partial\phi\bar{\partial}\phi
 -{\phi\over 4}\sqrt{g}R-\beta\bar{\partial}\gamma-\bar{\beta}
 \partial\bar{\gamma}\Big]~.
\end{equation}
The same $\phi_0$
dependence as in (\ref{index}) comes from the term
$\phi \sqrt{g} R$ 
 and a similar calculation in the above is possible.
However, in this approach, 
the full $SL(2)$ symmetry (\ref{fullsym}) appears subtle: 
in the full interacting theory, it is difficult to evaluate the Jacobian 
from (\ref{fullmeasure}) to (\ref{flatmeasure}) and Jacobians
of the type ${\cal D} (e^{\epsilon(\gamma)} \gamma)/{\cal D} {\gamma}$,
which are needed to check the invariance under (\ref{fullsym}).\footnote{
Strictly speaking, one needs to check that the regularization 
in calculating (\ref{chiralanom}) respects the $SL(2)$ symmetry.
However, our results indicate that the procedure in the above actually
respect it in total. 
}

\vspace*{3mm}
 Alternatively, the expression (\ref{freeint}) can be obtained
 by starting with the free theory with the action $S_{\rm free}$
 and a perturbation term $S_{\rm int}$.
 In this picture, all fields are free fields from the beginning
 and, for example, \ $\beta,\gamma (\bar{\beta},\bar{\gamma})$
 constitute a holomorphic 
(anti-holomorphic) bosonic ghost system. $\Phi_j^{\rm f}$ are 
 the primary fields with respect to the standard free field
$SL(2)$ currents:
\begin{equation}
  \hat{j}^- = \beta ~, \quad 
  \hat{j}^3 = \beta\gamma + b^{-2}\partial\phi ~, \quad 
  \hat{j}^+ = \beta\gamma^2 + 2b^{-2}\gamma\partial\phi + k\partial\gamma ~.
\end{equation} 
In fact, $\Phi_j^{\rm f}$ satisfy the OPEs (\ref{jPhi}) with $\hat{j}^A$
and have worldsheet conformal weight $h\equiv -b^2j(j+1)$.
%
One then finds that 
   the interaction $S_{\rm int}$ is made of a screening current
 $\beta \bar{\beta} e^{-2\phi}$ which has no singular OPEs with 
  $ \hat{j}^A $
 up to total derivatives.

 Generic correlators in this case are defined as follows:
\begin{equation}
  \vev{X[\phi,\beta,\gamma,\bar{\beta},\bar{\gamma}]}
  \equiv
  \int {\cal D}\phi{\cal D}\gamma{\cal D}\bar{\gamma}
       {\cal D}\beta{\cal D}\bar{\beta}
  \exp[-S_{\rm free}]\cdot X\exp[-S_{\rm int}] ~.
 \label{CorrFree1}
\end{equation}
Now the $SL(2)$ currents $\hat{j}^A$ are associated to
 the following symmetry of the path-integration:
\begin{eqnarray}
&& ~~~~~
  \int {\cal D}\phi'{\cal D}\gamma'{\cal D}\bar{\gamma}'
       {\cal D}\beta'{\cal D}\bar{\beta}'
  \exp[-S'_{\rm free}]
  =
  \int {\cal D}\phi{\cal D}\gamma{\cal D}\bar{\gamma}
       {\cal D}\beta{\cal D}\bar{\beta}
  \exp[-S_{\rm free}] ~,
\nonumber \\&&\hspace*{-6mm}
 \gamma' = \gamma+\epsilon ~,~~
 \beta'  = \beta - \epsilon'\beta
           - b^{-2}\epsilon''\partial\phi
           + \frac{k}{2} \partial \epsilon''~,~~
 \phi'   = \phi - \frac{1}{2}\epsilon' ~, 
\label{freesym}
\end{eqnarray}
where $\epsilon$ is as given in (\ref{fullsym}).\footnote{
Here, there is a subtlety again in evaluating  Jacobians such as  
${\cal D}( e^{\epsilon(\gamma)} \gamma)/{\cal D} {\gamma}$.
}  
However, the above transformation leaves $S_{\rm int}$ invariant only up to 
terms proportional to the equation of motion.
 Hence it is an on-shell or perturbative symmetry, but not the symmetry
 of the full theory based on the original action (\ref{S-PG}).

   Integrating over non-zero modes in (\ref{CorrFree1}), we obtain
\begin{equation}
  \vev{X} = \int d\phi_0 d\gamma_0 d\bar{\gamma}_0
            d^g\beta_0 d^g\bar{\beta}_0
            e^{2(1-g)\phi_0}
            \vev{X\exp[-S_{\rm int}]}^{\rm f}_{\phi_0,\gamma_0,
                               \bar{\gamma}_0,\beta_0,\bar{\beta}_0} ~,
\end{equation}
 where the bracket $\vev{A}^{\rm f}_{\phi_0,\cdots}$
 represents the Wick contraction of $A$ as before.
 For correlators of primaries on a sphere,
 we actually obtain the same expression as in (\ref{freeint})
 by substituting $X=\prod_{a=1}^N \Phi^{\rm f}_{j_a}(z_a,x_a)$
 and integrating over $\phi_0$.

   Although we can obtain the same expressions for correlators,
 the underlying symmetry (\ref{freesym}) 
  is  different
 from that of the full interacting theory as discussed above.
 Hence it is more appropriate to start from the full treatment
 if we try to analyze the original theory with the full symmetry 
 (\ref{fullsym})
  and regard the invariance of correlation functions
 as originating from the true symmetry of the Lagrangian.

   The procedure leading to (\ref{freeint}) may be an analog
 of the Liouville case discussed by Goulian and Li 
 \cite{Goulian-L} (see also \cite{Braaten-CGT,Gupta-TW,Bershadsky}).
 Since the $H_3^+$ theory is a little more complicated
 than  Liouville theory, we needed to take into account
 the renormalization in addition to the integral over $\phi_0$.

   Using the free field prescription obtained in this way and,
 in particular, the formula (\ref{freeint}),
 we would like to obtain the explicit forms of the correlators
 and analytically continue them in $j_a$ similarly to  
\cite{Dorn-O,Zamolodchikov-Z}. 
 Such a continuation may be justified along the same line
 as in the Liouville case \cite{Aoki-DH,Gervais}.
 Indeed, we will see that our correlators are in 
 complete agreement with the exact results obtained by
 other approaches \cite{Teschner1,Teschner2,Ishibashi-OS}.

   For the expression (\ref{freeint}) to make sense,
 $\sum_a j_a + 1$ should be a non-negative integer, but
 the prefactor $\Gamma(-\sum_aj_a - 1)$ is then divergent.
 Similar divergences appear also in Liouville theory.
 In that case, they arise inevitably if we define
 correlation functions as analytic functions of complex $j_a$
 \cite{Dorn-O,Zamolodchikov-Z}.
 The situation in our case seems similar and
 $\sum_a j_a + 1 \in {\bf Z}_{\geq 0}$ may be interpreted as
 a kind of ``mass-shell'' condition according to the Liouville case 
 \cite{DiFrancesco-K,Zamolodchikov-Z}.

 Also, the free field approach is usually taken to be valid as 
 $\phi \to \infty$, namely,
 near the boundary of $AdS_3$, because the interaction term
 $S_{\rm int}$ is vanishing there.
 The primaries $\Phi_j$ reduce to $\Phi_j^{\rm f}$ in that limit. 
 However, an important consequence of our argument is that
 the free field approach is more powerful as long as we consider
 the correlation functions of the primaries. 

 Finally, we would like to comment on the relationship to \cite{Ishibashi-OS}.
  Our new prescription differs from those in \cite{Gawedzki}
  and \cite{Ishibashi-OS} in which correlators are defined  by
  taking projections onto the $SL(2,{\bf C})$-invariant part by hand 
   and introducing 
  delta-functionals such as $\delta^2(e^{2\phi(z_0)}\gamma(z_0))$.
 Because of such functionals, $N$-point functions are seemingly 
 represented by $(N+1)$-point functions in an ordinary sense.
  On the other hand, in our formula, an $SL(2,{\bf C})$-invariant result
  (in the sense of space-time) is obtained by the integration over
  the zero-modes of $\phi,\gamma,\bar{\gamma}$, and the calculation
  is very much simplified. One can confirm  such simplifications in the 
   following examples of the two- and three-point functions.


\mysubsection{Two-Point Function}
 For the two-point function, the formula (\ref{freeint}) and the 
integration over $\gamma_0, \bar{\gamma}_0$ yield
\begin{eqnarray*}
\lefteqn{\vev{\Phi_{j_1}(z_1,x_1)\Phi_{j_2}(z_2,x_2)}} \\
    &=&
  \frac{\pi}{2}\Gamma(-j_1-j_2-1)
  \Delta(2j_1+1)\Delta(2j_2+1)\Delta(-j_1-j_2)\Gamma(j_1+j_2+2)^{-2}
  |x_{12}|^{2j_1+2j_2}
  \\ &&  ~~~\cdot
  \vev{|\gamma_{12}|^{2j_1+2j_2+2}
       e^{2j_1\phi(z_1)}e^{2j_2\phi(z_2)}
       \left[-S_{\rm int}\right]^{j_1+j_2+1}
       }^{\rm f}_{\phi_0=\gamma_0=\bar{\gamma}_0=0} ~.
\end{eqnarray*}
 Here $\gamma_{12}= \gamma(z_1)-\gamma(z_2)$, $x_{12}=x_1-x_2$
 and we have introduced $\Delta(x)\equiv \Gamma(x)/\Gamma(1-x)$.
 The integration over $\gamma_0$ can be carried out
 using the formula
\[
 \int d^2x |x|^{4j_1}|1-x|^{4j_2} =
 \pi \Delta(2j_1+1)\Delta(2j_2+1)\Delta(-2j_1-2j_2-1) ~,
\]
 and then we expanded a binomial
 $|\gamma_{12}-x_{12}|^{4j_1+4j_2+2}$ and picked up the
 relevant term.
 The remaining free CFT correlator is given by a
 Dotsenko-Fateev integral \cite{Dotsenko-F1}:
\begin{eqnarray*}
 && \frac{1}{\Gamma(j_1+j_2+2)^2}
  \vev{|\gamma_{12}|^{2j_1+2j_2+2}
       e^{2j_1\phi(z_1)}e^{2j_2\phi(z_2)}
       \left[\int\frac{d^2w}{k\pi}\beta\bar{\beta}e^{-2\phi}(w)
       \right]^{j_1+j_2+1}
       }^{\rm f}_{\phi_0=\gamma_0=\bar{\gamma}_0=0}
 \\ &&
 ~~ = ~~
  |z_{12}|^{2b^2j_1(j_1+1)+2b^2j_2(j_2+1)}
  K(j_1-\frac{1}{2b^2},~j_2-\frac{1}{2b^2},~0),
\end{eqnarray*}
 where
\begin{eqnarray}
 K(\alpha_1,\alpha_2,\alpha_3)
 &\equiv&
 \int \prod_{i=1}^n \frac{d^2y_i}{k\pi}
  |y_i|^{4b^2\alpha_1}|1-y_i|^{4b^2\alpha_2}
  \prod_{i<j}|y_i-y_j|^{-4b^2} \nonumber \\
 &=&
 \frac{[k^{-1}b^{-2b^2}\Delta(b^2)]^n
        \Upsilon[b]
        \Upsilon[-2\alpha_1b]
        \Upsilon[-2\alpha_2b]
        \Upsilon[-2\alpha_3b]}
       {\Gamma(n+1)
        \Upsilon[-(\sum\alpha_i+1)b]
        \Upsilon[-\alpha_{12}b]
        \Upsilon[-\alpha_{23}b]
        \Upsilon[-\alpha_{31}b]} ~,
\label{K}
\end{eqnarray}
\[
  n=\sum \alpha_i+1+b^{-2}  ~,~~
  \alpha_{12}= \alpha_1+\alpha_2-\alpha_3~,~~
  \alpha_{23}= \alpha_2+\alpha_3-\alpha_1~,~~
  \alpha_{13}= \alpha_1+\alpha_3-\alpha_2 ~.
\]
 Here the integral is expressed using the $\Upsilon$-function.
 The definition and some basic properties of $\Upsilon(x)$
 are found, e.g., in the appendix of \cite{Ishibashi-OS}.

   In calculating further, note that $K(\alpha_i)$ becomes
 delta-functional in the limit $\alpha_3\rightarrow 0$ with
 the support on the zeroes of the denominator.
 Analyzing in a similar way as in \cite{Teschner2,Ishibashi-OS},
 we see that the relevant zeroes of the denominator are
 at $j_1+j_2+1=0$ and at $j_1=j_2$. 
 Thus we obtain
\begin{eqnarray}
 \lefteqn{ \vev{\Phi_{j_1}(z_1,x_1)\Phi_{j_2}(z_2,x_2)}}
 ~~~~~~~~ \nonumber \\
 &=&
 |z_{12}|^{4b^2j_1(j_1+1)}[A(j_1)\delta^2(x_{12})i\delta(j_1+j_2+1)
                          +B(j_1)|x_{12}|^{4j_1} i\delta(j_1-j_2)] ~,
 \nonumber \\
 A(j) &=& -\frac{\pi^3}{(2j+1)^2} ~, \nonumber \\
 B(j) &=& b^2\pi^2[k^{-1}\Delta(b^2)]^{2j+1}\Delta[-b^2(2j+1)] ~.
\end{eqnarray}
 Comparing this with the result of \cite{Ishibashi-OS}, we see that
 they agree precisely up to an overall numerical factor.
 We also find an agreement with \cite{Teschner1,Teschner2}
 by appropriate changes of normalizations of the primaries.

\mysubsection{Three-Point Function}

   In calculating the three-point function,
 one has to make use of the $SL(2)$ symmetry (\ref{jPhi}), and   
  extract  the $x_a$-dependence
 in the following way:
\begin{eqnarray}
  \vev{\prod_{a=1}^3\Phi_{j_a}(z_a,x_a)}
 &=& \prod_{a<b}|x_{ab}|^{2j_{ab}}D(j_a,z_a) ~,\nonumber \\
  D(j_a,z_a) &=&
 \prod_{a<b}|x_{ab}|^{-2j_{ab}}
  \left.
  \vev{\prod_{a=1}^3\Phi_{j_a}(z_a,x_a)}
  \right|_{x_{1,2,3}=0,1,\infty} ~,
\end{eqnarray}
 where we have used the notation $j_{12}=j_1+j_2-j_3$, etc.
 The coefficient $D(j_a,z_a)$ can then be calculated
 using (\ref{freeint}).
 After separating the $z_a$-dependence by means of the worldsheet
 $SL(2,{\bf C})$ invariance, we find that the remaining part
 is again described by a Dotsenko-Fateev integral:
\begin{eqnarray}
  D(j_a,z_a) &=& \prod_{a<b}|z_{ab}|^{-2h_{ab}}\cdot
  \frac{\pi}{2}\Gamma(-\Sigma j_a -1)\Delta(2j_1+1)\Delta(2j_2+1)
  \Delta(-j_{12}) \nonumber \\ &&
  \cdot K(j_1-\frac{1}{2b^2},j_2-\frac{1}{2b^2},j_3) ~.
\end{eqnarray}
 Here we have used $h_{12} = h_1+h_2-h_3$, etc.
 Summarizing, the three-point function is given by
\begin{eqnarray}
 && ~~~~~
  \vev{\prod_{a=1}^3\Phi_{j_a}(z_a,x_a)}
  ~=~ D(j_a)\prod_{a<b}|z_{ab}|^{-2h_{ab}}|x_{ab}|^{2j_{ab}} ~,
 \nonumber \\&&
 D(j_a) ~=~
  \frac{b^2\pi}{2}
  \frac{[k^{-1}b^{-2b^2}\Delta(b^2)]^{\Sigma j_a+1}
        \Upsilon[b]\Upsilon[-2j_1b]\Upsilon[-2j_2b]\Upsilon[-2j_3b]}
       {\Upsilon[-(\Sigma j_a+1)b]\Upsilon[-j_{12}b]\Upsilon[-j_{13}b]
        \Upsilon[-j_{23}b]} ~.
\end{eqnarray}
 This is again in precise agreement with the known results.

   Before concluding this section, we would like to note
 that our two- and three-point functions are 
 consistent with the following symmetry of primary fields:
\begin{eqnarray}
  \Phi_j(z,x) &=& R(j) \int d^2y |x-y|^{4j}\Phi_{-j-1}(z,y)~, \nonumber \\
  R(j) &=& -\frac{(2j+1)^2b^2}{\pi}\Delta[-(2j+1)b^2]
           \left[k^{-1}\Delta(b^2)\right]^{2j+1} ~.
\label{REFLECT}
\end{eqnarray}
This is understood
 as a non-trivial check that the procedure in subsection 3.1 respected 
 the $SL(2)$ symmetry.

    It is straightforward to write down an integral formula
 for $N$-point functions of primary fields. We will give the explicit
 expression in the $N=4$ case in the next section.



\mysection{Solving Knizhnik-Zamolodchikov equation}

 To see that our prescription works also for four-point functions,
 we would like to obtain some of them from a different approach:
 by solving the Knizhnik-Zamolodchikov(KZ) equation.
  We will find that
 some solutions obtained in this way are also calculated correctly
 and easily from our prescription in section 3.

 For generic correlators of primaries, the KZ equation is given by
\begin{equation}
 \left[ \partial_{z_a}- b^2 
 \sum_{b(\neq a)}z_{ab}^{-1}{\cal L}_{ab}
 \right] \vev{\prod_c\Phi_{j_c}(z_c,x_c)} = 0
 ~,
\end{equation}
\[
 {\cal L}_{ab}
    =  x_{ab}^2\partial_{x_a}\partial_{x_b}
       -2x_{ab}(j_a\partial_{x_b}-j_b\partial_{x_a})-2j_aj_b ~.
\]
  In the case of four-point functions, 
   the worldsheet and space-time
 $SL(2,{\bf C})$ invariances determine their form
 up to an arbitrary function of cross ratios:
\begin{equation}
  \vev{\prod_{a=1}^4\Phi_{j_a}(z_a,x_a)}
 = \prod_{a<b}^4|z_{ab}|^{2b^2\mu_{ab}}|x_{ab}|^{2\lambda_{ab}}F(z,x) ~,
\end{equation}
\begin{equation}
 z \equiv \frac{z_{41}z_{23}}{z_{43}z_{21}}~,~~~
 x \equiv \frac{x_{41}x_{23}}{x_{43}x_{21}}~.
\end{equation}
 If we choose $\lambda_{ab}$ and $\mu_{ab}$ in the following way,
\begin{equation}
\begin{array}{rcl}
 \lambda_{12} &=& j_1+j_2-j_3+j_4 ~,\\
 \lambda_{13} &=& j_1-j_2+j_3-j_4 ~,\\
 \lambda_{14} &=& 0 ~,\\
 \lambda_{23} &=& -j_1+j_2+j_3-j_4 ~,\\
 \lambda_{24} &=& 0 ~,\\
 \lambda_{34} &=& 2j_4 ~,
\end{array}
~~
\begin{array}{rcl}
 \mu_{12} &=& \Delta_1+\Delta_2-\Delta_3+\Delta_4+2j_1j_4+2j_2j_4 ~,\\
 \mu_{13} &=& \Delta_1-\Delta_2+\Delta_3-\Delta_4-2j_2j_4 ~,\\
 \mu_{14} &=& -2j_1j_4 ~,\\
 \mu_{23} &=& -\Delta_1+\Delta_2+\Delta_3-\Delta_4-2j_1j_4 ~,\\
 \mu_{24} &=& -2j_2j_4 ~,\\
 \mu_{34} &=& 2\Delta_4+2j_1j_4+2j_2j_4 ~,
\end{array}
\end{equation}
 with $\Delta_a = j_a(j_a+1)$, the KZ equation for $F(z,x)$ becomes
\begin{eqnarray}
  0 &=&
  \left[\frac{1}{b^2}\frac{\partial}{\partial z}
       +\frac{xP_0}{z}+\frac{(1-x)P_1}{z-1}\right]F ~,
 \label{K-Z}\\
  P_i &=& x(1-x)\partial_x^2
        +\left[\gamma_i-(1+\alpha+\beta)x\right]\partial_x-\alpha\beta ~,
 \nonumber
\end{eqnarray}
\[
 \alpha   = -2j_4            ~,~~~
 \beta    = -j_1-j_2+j_3-j_4 ~,~~~
 \gamma_0 = -2j_1-2j_4       ~,~~~
 \gamma_1 = 1-j_1+j_2+j_3-j_4 ~.
\]
 We see that $P_i$ in the above are nothing but the
 hypergeometric differentials.
 Hence it is expected that, under certain conditions on $j_a$,
 the solution can be written down explicitly using
 hypergeometric functions.
 In the following we will give some examples in which the
 KZ equation is solved rather easily.

\mysubsection{$\sum_a j_a = -1$}
 The simplest example is the case $\sum_a j_a =-1$.
 Since $\gamma_0=\gamma_1$ and the two hypergeometric
 differentials coincide in this case, one can find
 a solution with $F(z,x)$ independent of $z$.
 The four-point function is therefore given by
\begin{eqnarray}
 \vev{\prod_{a=1}^4\Phi_{j_a}(z_a,x_a)} &=&
 |x_{12}|^{2(j_1+j_2-j_3+j_4)}
 |x_{13}|^{2(j_1-j_2+j_3-j_4)}
 |x_{23}|^{2(-j_1+j_2+j_3-j_4)}
 |x_{34}|^{4j_4} \nonumber \\ && \cdot
 \prod_{a<b}^4|z_{ab}|^{-4b^2j_aj_b}\cdot
 f(x) ~,
\end{eqnarray}
 where $f(x)$ satisfies the hypergeometric differential equation
 $P_0f(x)=0$.
 Recalling that $f(x)$ must be real and single-valued
 if $j_a$ are all real, one obtains
\begin{eqnarray}
 f(x)&=& C\int d^2t |t|^{2\beta-2}|1-t|^{2\gamma_0-2\beta-2}|1-tx|^{-2\alpha}
 \nonumber \\
     &=& C\int d^2t |t|^{4j_3}|1-t|^{4j_2}|1-tx|^{4j_4} ~,
\end{eqnarray}
 where $C$ is a constant.
 Actually this $f(x)$ is understood as a naive ``square'', i.e.,
 monodromy-invariant combination, 
 of the hypergeometric function represented as a contour integral.

   Making a change of the integration variable from $t$ to $x_0$
 defined by
\[
 t=\frac{x_{03}x_{21}}{x_{01}x_{23}} ~,
\]
 the solution can be recast in a form which is manifestly symmetric
 in four vertices:
\begin{equation}
 \vev{\prod_{a=1}^4\Phi_{j_a}(z_a,x_a)} =
 C(j_a)\prod_{a<b}^4|z_{ab}|^{-4b^2j_aj_b}\cdot
 \int d^2x_0 \prod_a |x_{0a}|^{4j_a} ~.
\label{SOL1}
\end{equation}

\mysubsection{$\sum_a j_a = 0$}

   The next simplest is the case $\sum_a j_a = 0$ or $\gamma_1=\gamma_0+1$.
 In this case the two hypergeometric functions
 $F(\alpha,\beta,\gamma_i;x)$ have a special property:
 the differentials $xP_0$ and $(1-x)P_1$ are linearly realized
 on them.
 This is due to the following recursion relation of
 hypergeometric functions:
\begin{eqnarray}
  \gamma(1-x)\frac{\partial}{\partial x}F(\alpha,\beta,\gamma;x)
 &=& (\gamma-\alpha)(\gamma-\beta)F(\alpha,\beta,\gamma+1;x)
     +\gamma(\alpha+\beta-\gamma)F(\alpha,\beta,\gamma;x) ~,\nonumber \\
  x\frac{\partial}{\partial x}F(\alpha,\beta,\gamma+1;x)
 &=& \gamma\left\{F(\alpha,\beta,\gamma;x)
      -F(\alpha,\beta,\gamma+1;x) \right\} ~.
 \label{REC}
\end{eqnarray}
 Therefore the solution to the KZ equation can be found as a
 linear combination of the two hypergeometric functions
 $F(\alpha,\beta,\gamma_i;x)$,
\begin{equation}
  F(z,x) \sim
  (z-1)g_0(z)F(\alpha,\beta,\gamma_0;x)
  +   zg_1(z)F(\alpha,\beta,\gamma_1;x)~,
\end{equation}
 or of another solutions to $P_if_i(x)=0$.
 Here ``$\sim$'' means that we do not care about the dependence on
 $\bar{z}$ and $\bar{x}$ for the time being.

   Putting the above ansatz into the KZ equation,
 we obtain
\begin{eqnarray*}
  \frac{\partial}{b^2\partial z}[(z-1)g_0]  &=&
  ~~\gamma_0g_1 - (\alpha+\beta-\gamma_0)g_0   ~,\\
  \frac{\partial}{b^2\partial z}[   z g_1]  &=&
  -\gamma_0g_1 - \frac{(\gamma_0-\alpha)(\gamma_0-\beta)}{\gamma_0}g_0 ~.
\end{eqnarray*}
 The previous recursion relation is utilized here again,
 and we obtain a solution to the KZ equation which has the following form:
\begin{eqnarray}
  F(z,x) ~\sim~
  \gamma_0(1+b^2\gamma_0)(1-z)
  F(1+b^2\alpha,1+b^2\beta,1+b^2\gamma_0;z)
  F(\alpha,\beta,\gamma_0;x)
   && \nonumber \\
 +b^2(\gamma_0-\alpha)(\gamma_0-\beta)z
  F(1+b^2\alpha,1+b^2\beta,2+b^2\gamma_0;z)
  F(\alpha,\beta,\gamma_0+1;x)  &&
\end{eqnarray}
 Since there are two solutions to each hypergeometric
 differential equation, we have three other independent
 solutions to the KZ equation which have a similar form.

   Now we rewrite the above solution 
 using the contour integral expression for the hypergeometric function,
 so that the permutation symmetry among four vertices becomes manifest.
 Writing $F(z,x)$ as
\begin{eqnarray}
  F(z,x) &=& \gamma_0(1+b^2\gamma_0)(1-z)G_0(z)F_0(x)
            +b^2(\gamma_0-\alpha)(\gamma_0-\beta)zG_1(z)F_1(x)
   ~, \nonumber \\
  G_0(z) &=& \frac{1}{1-e^{2\pi i(b^2\beta+1)}}
             \frac{\Gamma(b^2\gamma_0+1)}
                  {\Gamma(b^2\beta+1)\Gamma(b^2(\gamma_0-\beta))}
             \oint dt t^{b^2\beta}(1-t)^{b^2(\gamma_0-\beta)-1}
                      (1-tx)^{-b^2\alpha-1}
   ~, \nonumber \\
  G_1(z) &=& \frac{1}{1-e^{2\pi i(b^2\beta+1)}}
             \frac{\Gamma(b^2\gamma_0+2)}
                  {\Gamma(b^2\beta+1)\Gamma(b^2(\gamma_0-\beta)+1)}
             \oint dt t^{b^2\beta}(1-t)^{b^2(\gamma_0-\beta)}
                  (1-tx)^{-b^2\alpha-1}
   , \nonumber \\
  F_0(x) &=& \frac{1}{1-e^{2\pi i\beta}}
             \frac{\Gamma(\gamma_0)}{\Gamma(\beta)\Gamma(\gamma_0-\beta)}
             \oint ds s^{\beta-1}(1-s)^{\gamma_0-\beta-1}(1-sx)^{-\alpha}
   ~, \nonumber \\
  F_1(x) &=& \frac{1}{1-e^{2\pi i\beta}}
             \frac{\Gamma(\gamma_0+1)}
                  {\Gamma(\beta)\Gamma(\gamma_0-\beta+1)}
             \oint ds s^{\beta-1}(1-s)^{\gamma_0-\beta}(1-sx)^{-\alpha}
\label{CONTOUR}
\end{eqnarray}
 and making the change of variables from $s,t$ to $x_0, z_0$
 which are defined by
\[
   s= \frac{x_{03}x_{21}}{x_{01}x_{23}} ~,~~
   t= \frac{z_{03}z_{21}}{z_{01}z_{23}} ~,
\]
 we can rewrite $F(z,x)$ in the following form:
\begin{equation}
  \prod_{a<b}^4
  {z_{ab}}^{b^2\mu_{ab}}{x_{ab}}^{\lambda_{ab}}F(z,x) \sim
  \prod_{a<b}^4
  {z_{ab}}^{-2b^2j_aj_b}
  \oint dz_0dx_0 \prod_a z_{0a}^{2b^2j_a}x_{0a}^{2j_a}
  \cdot\sum_a\frac{j_a}{z_{0a}x_{0a}} ~.
\end{equation}
 The above expression is obviously symmetric in four vertices,
 and by taking its naive square we obtain the four-point function:
\begin{equation}
  \vev{\prod_{a=1}^{4}\Phi_{j_a}(z_a,x_a)}
 = C(j_a)
  \prod_{a<b}|z_{ab}|^{-4b^2j_aj_b}
  \int d^2z_0d^2x_0 \prod_a |z_{0a}|^{4b^2j_a}|x_{0a}|^{4j_a}
  \cdot\left|\sum_a\frac{j_a}{z_{0a}x_{0a}}\right|^2 ~.
\label{SOL2}
\end{equation}

   The other solutions to the KZ equation can be obtained
 by starting from (\ref{CONTOUR}) with $\alpha$ and $\beta$
 in $F_{0,1}(x)$ exchanged each other.
 Proceeding in a similar way we obtain
\begin{eqnarray*}
  z_{ab}^{b^2\mu_{ab}}x_{ab}^{\lambda_{ab}}F(z,x)
 &\sim& z_{ab}^{-2b^2j_aj_b}\oint dz_0dx_0 \prod_a z_{0a}^{2b^2j_a}x_{0a}^{2j_a}
 \\ && ~~~~~\cdot
  \left\{ (x_{01}^{-1}-x_{04}^{-1})(z_{02}^{-1}-z_{04}^{-1})
         -(x_{02}^{-1}-x_{04}^{-1})(z_{01}^{-1}-z_{04}^{-1})
  \right\} ~.
\end{eqnarray*}
 By exchanging four vertices we obtain different solutions,
 but the number of independent solutions is three.
 By summing over their squares with an appropriate weight,
 it might be possible to construct a quantity
 satisfying consistency conditions such as the single-valuedness,
 symmetry between four vertices, etc.
 However, it cannot reduce to the known three-point function
 of the $H_3^+$ WZW model in the limit $j_4\rightarrow 0$
 as one can see by a comparison with
 the free field result in section 3. Note that, since our  KZ equation
is a partial differential equation, there exist large degrees of 
freedom of the solutions.

\mysubsection{Comparison with Free Field Approach}

   By simple calculation we see that  (\ref{freeint})
 with $N=4$ gives the solutions to the KZ equation,
 (\ref{SOL1}) and (\ref{SOL2}), up to factors.
 In this identification, the integration variables $x_0$ and $z_0$ in 
 (\ref{SOL1}) and (\ref{SOL2}) correspond to the zero-mode
 of $\gamma$ and the position of the screening current
 $\beta\bar{\beta} e^{-2\phi}$, respectively.  

   We can write down the integral formula for more generic
 four-point functions.
 If $\sum_a j_a + 1$ is a non-negative integer,
 the four-point function is expressed in terms of
 a free field correlator with $\sum_a j_a+1$ insertions of $S_{\rm int}$:
\begin{eqnarray}
&&\hspace*{-6mm}
 \vev{\prod_{a=1}^4\Phi_{j_a}(z_a,x_a)} ~=~
 \frac{1}{2}\Gamma(-\Sigma j_a-1)(-k\pi)^{-\Sigma j_a-1}
 \prod_{a<b}|x_{ab}|^{2\lambda_{ab}}|z_{ab}|^{2b^2\mu_{ab}}|f(z,x)|^2 ~,
 \nonumber \\
&&\hspace*{-6mm}
 f(z,x) ~=~
 \frac{\Gamma(j_1-j_2+j_3-j_4+1)}{\Gamma(-2j_2)\Gamma(-2j_4)}
 \int d\gamma_0 \cdot \gamma_0^{2j_1} \nonumber \\
&&
 \cdot \int_0^1 dsdt\delta(s+t-1) s^{-2j_2-1}t^{-2j_4-1}
 (\gamma_0-s-tx)^{-j_1+j_2-j_3+j_4-1}
 \nonumber \\
&&
 \cdot \int \prod_{i=1}^{\Sigma j_a+1}dy_i
 \left(\frac{s}{y_i-1}+\frac{tz}{y_i-z}\right)
 y_i^{2b^2j_1-1}(y_i-1)^{2b^2j_2}(y_i-z)^{2b^2j_4}
 \prod_{i<j}y_{ij}^{-2b^2} ~.
\end{eqnarray}

   As a remark, we note that our new formula with an additional
 integration over the zero-mode of $\gamma$ is related to the
 well-studied case $j_1+j_2-j_3+j_4 \in {\bf Z}_{\ge 0}$ in the literature.
 In this case, using the free field realization of the $SL(2)$ current, 
 the solutions to the KZ equation for $F(z,x)$ have been given
 by the free field correlators with the insertion of $j_1+j_2-j_3+j_4$
 screening charges \cite{Furlan-GPP,Petersen-RY,Saraikin}. 
 Note that, when $j_1+j_2-j_3+j_4 \in {\bf Z}_{\ge 0}$, the KZ equation
 can be solved rather easily, because $\beta$ in (\ref{K-Z})
 is a negative integer and the hypergeometric function then reduces
 to a polynomial of $x$.
 Those solutions and ours may be 
  connected  through the relation
 (\ref{REFLECT}) between $\Phi_{j_3}$ and $\Phi_{-j_3-1}$.
 For example, the polynomials of $x$ in the former are converted 
 to the hypergeometric functions in the latter 
by the $y$-integral in (\ref{REFLECT}).   
 
\mysection{Conclusions}

   In this paper, we have argued that the 
 calculation of  the correlators in the $H_3^+$ WZW model 
 is reduced to that of a free theory. 
 The point is the integration over the zero-mode of $\phi$
 and the renormalization.
 
   Since the free field picture used in the calculation has been deduced 
 from the full interacting theory, the results are expected to be exact.
 Indeed, we have found a precise agreement with the known exact results
 for the two- and three-point functions.
 Moreover, our prescription allows us to calculate
 higher point correlation functions including numerical factors, 
 at least under some conditions.
 We have actually seen that the four-point functions which are
 obtained by solving the KZ equation can also be reproduced
 easily in our formalism.

   In comparing the solutions to the KZ equation with our free field
 results, we identify one of the integration variables in the
 former with the zero-mode of $\gamma$ in the latter.
 This means that we have to deal with the zero-modes of the fields
 appropriately in calculating correlators even if we use
 the free field picture.

  Admittedly, we have not yet succeeded in extracting the
 results directly relevant to the string theory on $AdS_3$.
 However, we believe that our argument here 
 made a step toward a better understanding. Thus, let us discuss some possible 
 directions.
     First, one advantage of using the free field picture may be that we can
 construct various vertices easily.
 For example, we need some free field realization of the current algebra
 to construct vertices for
 spectral-flowed representations \cite{Maldacena-O}.
 The correlators containing such vertices may also be calculated
 in our approach.

 Second, it is interesting to study the singular behavior
 of the four-point function when $z$ approaches $x$ in $F(z,x)$.
 In some recent work, such a singularity is identified with the singularity
 involving a ``long string'' with a unit winding number \cite{talk}.
  To investigate the four-point functions further, 
  it  may be useful to use the relationship between
   the four-point function in the $SL(2)$ case and the five-point
   function in the minimal models 
  \cite{Zamolodchikov-F,Furlan-GPP,Petersen-RY}

  Regarding the AdS/CFT correspondence, the Virasoro
 central charge of the space-time CFT might be calculable
 from our approach.
 It would be interesting to obtain its precise value
 and consider why it has to be quantized.
 Though the quantization is considered to be a non-perturbative
 effect as has been discussed in \cite{Kutasov-S},
 we might be able to explain it from our approach.

   Finally, we would like to note the similarity between
 our formalism and that for Liouville theory.
 In particular, in Liouville theory or the two-dimensional string theory, 
 the $N$-point correlators can be obtained when the moduli
 are integrated out \cite{DiFrancesco-K}. 
 In our case, such correlators correspond
 to the correlators of the ``space-time'' CFT. 
 Our discussion in this paper suggests that the correlators of the 
 space-time CFT may be analyzed using the techniques developed for 
 Liouville theory.  
  
\vskip 12mm
\begin{center}{\sc Acknowledgments}\end{center}
\par\smallskip

   We would like to thank N. Ishibashi for useful discussions
 and correspondences.
 We are also grateful to the organizers
 of ``Summer Institute 2000'' at Yamanashi, Japan, where
 a part of this work has been carried out.
 The works of K.H. and K.O. were supported in part
 by JSPS Research Fellowships for Young Scientists.
 The work of Y.S. was supported in part by Grant-in-Aid
 for Scientific Research on Priority Area 707 from
 the Ministry of Education, Science and Culture in Japan.

\def\thebibliography#1{\list
 {[\arabic{enumi}]}{\settowidth\labelwidth{[#1]}\leftmargin\labelwidth
  \advance\leftmargin\labelsep
  \usecounter{enumi}}
  \def\newblock{\hskip .11em plus .33em minus .07em}
  \sloppy\clubpenalty4000\widowpenalty4000
  \sfcode`\.=1000\relax}
 \let\endthebibliography=\endlist
\vskip 10ex
\begin{center}
 {\sc References}
\end{center}
\par


\begin{thebibliography}{999}
\parskip=-3pt


\bibitem{Maldacena} J. Maldacena,
   ``{\sl The large N limit of superconformal field theories
          and supergravity,}''
   Adv. Theor. Math. Phys. {\bf 2}, 231 (1998),
   {\tt hep-th/9711200}.

\bibitem{Gubser-KP} S.S. Gubser, I.R. Klebanov and A.M. Polyakov,
   ``{\sl Gauge theory correlators from non-critical string theory,}''
   Phys. Lett.  {\bf B428}, 105 (1998),
   {\tt hep-th/9802109}.

\bibitem{Witten} E. Witten,
   ``{\sl Anti-de Sitter space and holography,}''
   Adv. Theor. Math. Phys. {\bf 2}, 253 (1998),
   {\tt hep-th/9802150}.


\bibitem{Giveon-KS} A. Giveon, D. Kutasov and N. Seiberg,
  ``{\sl Comments on string theory on $AdS_3$,}''
   Adv. Theor. Math. Phys. {\bf 2}, 733 (1998),
  {\tt hep-th/9806194}.

\bibitem{deBoer-ORT} J. de Boer, H. Ooguri, H. Robins and
                            J. Tannenhauser,
  ``{\sl String theory on $AdS_3$,}''
   JHEP {\bf 9812}, 026 (1998),
   {\tt hep-th/9812046}.

\bibitem{Kutasov-S} D. Kutasov and N. Seiberg,
   ``{\sl More comments on string theory on $AdS_3$,}''
   JHEP {\bf 9904}, 008 (1999),
   {\tt hep-th/9903219}.


\bibitem{Bars-DM} I. Bars, C. Deliduman and D. Minic,
   ``{\sl String theory on $AdS_3$ revisited,}'' \\
   {\tt hep-th/9907087}.

\bibitem{Petropoulos} P.M. Petropoulos,
   ``{\sl String theory on $AdS_3$: some open questions,}''
   {\tt hep-th/9908189}.

\bibitem{Giribet-N1} G. Giribet and C. Nunez,
   ``{\sl Interacting strings on $AdS_3$,}''
   JHEP {\bf 9911}, 031 (1999),
   {\tt hep-th/9909149}.

\bibitem{Maldacena-O} J. Maldacena and H. Ooguri,
   ``{\sl Strings in $AdS_3$ and $SL(2,\bf{R})$ WZW model: I,}''
   {\tt hep-th/0001053}.

\bibitem{Kato-S} A. Kato and Y. Satoh,
   ``{\sl Modular invariance of string theory on $AdS_3$,}''
   Phys. Lett.  {\bf B486}, 306 (2000),
   {\tt hep-th/0001063}.

\bibitem{Larsen-S} A.L. Larsen and N. Sanchez,
   ``{\sl Quantum coherent string states in $AdS_3$ and
          $SL(2,{\bf R})$ WZWN model,}''
   Phys. Rev. {\bf D62}, 046003 (2000),
   {\tt hep-th/0001180}.

\bibitem{Pesando} I. Pesando,
   ``{\sl Some remarks on the free fields realization
          of the bosonic string on $AdS_3$,}''
   {\tt hep-th/0003036}.

\bibitem{Hikida-HS} Y. Hikida, K. Hosomichi and Y. Sugawara,
   ``{\sl String theory on $AdS_3$ as discrete light-cone
          Liouville theory,}''
  Nucl. Phys. {\bf B589}, 134 (2000), {\tt hep-th/0005065}.

\bibitem{Ishibashi-OS} N. Ishibashi, K. Okuyama and Y. Satoh,
   ``{\sl Path integral approach to string theory on $AdS_3$},''
   Nucl. Phys. {\bf B588}, 149 (2000), 
   {\tt hep-th/0005152}.

\bibitem{Maldacena-OS} J. Maldacena, H. Ooguri and J. Son,
   ``{\sl Strings in $AdS_3$ and the $SL(2,{\bf R})$ WZW model.
          II: Euclidean black hole,}''
   {\tt hep-th/0005183}.

\bibitem{Giribet-N2} G. Giribet and C. Nunez,
   ``{\sl Aspects of the free field description of
          string theory on $AdS_3$,}''
   JHEP {\bf 0006}, 033 (2000),
   {\tt hep-th/0006070}.


\bibitem{Haba} Z. Haba,
   ``{\sl Correlation functions of sigma fields with values
          in a hyperbolic space,}''
   Int. J. Mod. Phys. {\bf A4}, 267 (1989).

\bibitem{Gawedzki} K. Gaw{\c e}dzki,
   ``{\sl Quadrature of conformal field theories}'',
   Nucl. Phys. {\bf B328} (1989) 733; \\
   ``{\sl Non-compact WZW conformal field theories}'',
   NATO ASI: Cargese 1991: 0247-274,
   {\tt hep-th/9110076}.


\bibitem{Bernard-F}
   D. Bernard and G. Felder,
 ``{\sl Fock representations and BRST cohomology 
  in $SL(2)$ current algebra,}''
   Commun. Math. Phys.  {\bf 127}, 145 (1990).

\bibitem{Dotsenko}
   V.S.  Dotsenko,
 ``{\sl The free field representation of The $SU(2)$ 
  conformal field theory,}''
   Nucl. Phys.  {\bf B338}, 747 (1990); \\
   ``{\sl Solving the $SU(2)$ conformal field theory with the Wakimoto 
  free field representation,}''
   Nucl. Phys.   {\bf B358}, 547 (1991).

\bibitem{Gerasimov-MOMS} A. Gerasimov, A. Morozov, M. Olshanetsky,
                         A. Marshakov and S. Shatashvili,
   ``{\sl Wess-Zumino-Witten model as a theory of free fields,}''
   Int. J. Mod. Phys. {\bf A5}, 2495 (1990).

\bibitem{Furlan-GPP} P. Furlan, A.C. Ganchev, R. Paunov and V.B. Petkova,
  {\sl ``Reduction of the rational spin $sl(2,C)$ WZNW conformal theory,}''
  Phys.  Lett.  {\bf B267}, 63 (1991); \\
  ``{\sl Solutions of the Knizhnik-Zamolodchikov equation 
          with rational isospins and the reduction to the minimal models,}''
    Nucl. Phys. {\bf B394}, 665 (1993), {\tt hep-th/9201080}.

\bibitem{Andreev} O. Andreev,
   ``{\sl Operator algebra of the $SL(2)$ conformal field theories,}''
   Phys. Lett. {\bf B363}, 166 (1995),
   {\tt hep-th/9504082}.

\bibitem{Petersen-RY} J.L. Petersen, J. Rasmussen and M. Yu,
   ``{\sl Conformal blocks for admissible representations
          in $SL(2)$ current algebra,}''
   Nucl. Phys. {\bf B457}, 309 (1995),
   {\tt hep-th/9504127}; \\
%
   ``{\sl Hamiltonian reduction of SL(2) theories
          at the level of correlators,}''
    Nucl. Phys.  {\bf B457}, 343 (1995), {\tt hep-th/9506180}.

\bibitem{Andreev2} O. Andreev,
   ``{\sl On affine Lie superalgebras, $AdS_3/CFT$ correspondence
          and world-sheets for world-sheets,}''
    Nucl. Phys.  {\bf B552}, 169 (1999), {\tt hep-th/9901118}; \\
   ``{\sl Unitary representations of some infinite dimensional
          Lie algebras motivated by string theory on $AdS_3$,}''
    Nucl. Phys.  {\bf B561}, 413 (1999)  {\tt hep-th/9905002}.

\bibitem{Saraikin} K. Saraikin,
  ``{\sl Conformal blocks and correlators in WZNW model. I: Genus zero,}''
  {\tt hep-th/9912042}.


\bibitem{Teschner1} J. Teschner,
   ``{\sl On structure constants and fusion rules
          in the $SL(2,{\bf C})/SU(2)$ WZNW model,}''
   Nucl. Phys. {\bf B546}, 390 (1999),
   {\tt hep-th/9712256}; \\
   ``{\sl The mini-superspace limit of the $SL(2,{\bf C})/SU(2)$
          WZNW model,}''
   Nucl. Phys. {\bf B546}, 369 (1999),
   {\tt hep-th/9712258}.

\bibitem{Teschner2} J. Teschner,
   ``{\sl Operator product expansion and factorization
          in the $H_3^+$ WZNW model,}''
   Nucl. Phys. {\bf B571}, 555 (2000),
   {\tt hep-th/9906215}.

\bibitem{Kogan-LS}
I.I. Kogan, A. Lewis and O.A. Soloviev,
``{\sl Gauge Dressing of 2D Field Theories},''
Int. J. Mod. Phys. {\bf A12}, 2425 (1997), {\tt hep-th/9607048}; \\ 
``{\sl Knizhnik-Zamolodchikov-type equations for gauged WZNW
models},''
Int. J. Mod. Phys. {\bf A13},  1345 (1998), {\tt hep-th/9703028}.
%
\bibitem{Bhaseen-KSTT}
M.J. Bhaseen, I.I. Kogan, O.A. Soloviev, N. Taniguchi and A. Tsvelik,
``{\sl Towards a Field Theory of the Plateau Transition},''
Nucl. Phys. {\bf B580}, 688 (2000), {\tt cond-mat/9912060}.
%
\bibitem{Kogan-T}
I.I. Kogan and A.M. Tsvelik,
``{\sl Logarithmic Operators in the Theory of Plateau Transition},''
Mod. Phys. Lett. {\bf A15}, 931 (2000), {\tt hep-th/9912143}.
%
\bibitem{Nichols-S} A. Nichols and Sanjay,
   ``{\sl Logarithmic operators in the $SL(2,{\bf R})$ WZNW model,}''
   {\tt hep-th/0007007}.

\bibitem{Lewis}
A. Lewis,
``{\sl Logarithmic CFT on the Boundary and the World-Sheet},''
{\tt hep-th/0009096}.

\bibitem{Goulian-L} M. Goulian and M. Li,
   ``{\sl Correlation functions in Liouville theory,}''
   Phys. Rev. Lett. {\bf 66}, 2051 (1991).


\bibitem{Braaten-CGT} E.~Braaten, T.~Curtright, G.~Ghandour and C.~Thorn,
``{\sl Nonperturbative weak coupling analysis of 
 the Liouville quantum field theory,}''
  Phys. Rev. Lett.  {\bf 51}, 19 (1983).

\bibitem{Gupta-TW}  A.~Gupta, S.~P.~Trivedi and M.~B.~Wise,
 ``{\sl Random Surfaces In Conformal Gauge,}''
  Nucl. Phys.  {\bf B340}, 475 (1990).

\bibitem{Bershadsky}
M.~Bershadsky and D.~Kutasov,
``{\sl Comment on gauged WZW theory},''
Phys.\ Lett.\  {\bf B266}, 345 (1991).


\bibitem{Dorn-O} H. Dorn and H.J. Otto,
   ``{\sl Two and three point functions in Liouville theory,}''
   Nucl. Phys. {\bf B429}, 375 (1994),
   {\tt hep-th/9403141}.

\bibitem{Zamolodchikov-Z} A.B. Zamolodchikov and Al.B. Zamolodchikov,
   ``{\sl Structure constants and conformal bootstrap
          in Liouville field theory}'',
   Nucl. Phys. {\bf B477}, 577 (1996),
   {\tt hep-th/9506136}.



\bibitem{Aoki-DH} K.~Aoki and E.~D'Hoker,
``{\sl On the Liouville approach to correlation functions 
  for 2-D quantum gravity,}''
  Mod. Phys. Lett. {\bf A7}, 235 (1992), 
 {\tt hep-th/9109024}.

\bibitem{Gervais} J.~Gervais,
 ``{\sl Gravity - matter couplings from Liouville theory,}''
   Nucl. Phys.  {\bf B391}, 287 (1993),
 {\tt hep-th/9205034}.



\bibitem{DiFrancesco-K} P. Di Francesco and D. Kutasov,
   ``{\sl World sheet and space-time physics in two-dimensional
          (super)string theory,}''
   Nucl. Phys. {\bf B375}, 119 (1992),
   {\tt hep-th/9109005}.


\bibitem{Dotsenko-F1} V.S. Dotsenko and V.A. Fateev,
   ``{\sl Conformal algebra and multipoint correlation functions
          in 2D statistical models,}''
   Nucl. Phys. {\bf B240}, 312 (1984); \\
%
   ``{\sl Four point correlation functions and the operator algebra
          in the two-dimensional conformal invariant theories with
          the central charge $c < 1$,}''
   Nucl. Phys. {\bf B251}, 691 (1985).



\bibitem{Zamolodchikov-F} A.B. Zamolodchikov and V.A. Fateev,
   ``{\sl Operator algebra and correlation functions
          in the two-dimensional $SU(2)\times SU(2)$
          chiral Wess-Zumino model,}''
   Sov. J. Nucl. Phys. {\bf 43}, 657 (1986).

\bibitem{talk} H. Ooguri,
    talk at Strings 2000, \\ 
    { http://feynman.physics.lsa.umich.edu/cgi-bin/s2ktalk.cgi?ooguri}



\end{thebibliography}
\end{document}